\documentclass[12pt]{iopart}
\usepackage{graphicx}
\usepackage{epsf} 
\usepackage{dcolumn}
\usepackage{bm}
\usepackage{amssymb}
\usepackage{amsbsy}
\usepackage{graphicx,color,ifthen}
\usepackage{pstricks,pst-node,pst-text,pst-3d,color}

\usepackage{iopams}

\begin{document}

\title[]{One-dimensional pattern of Au nanodots by ion-beam-sputtering: formation and mechanism}

\author{J-H Kim$^1$, N-B Ha$^1$, J-S Kim$^1$, M Joe$^2$, K-R Lee$^2$ and R. Cuerno$^3$}

\address{$^1$ Department of Physics, Sook-Myung Women's University, Seoul 140-742, Korea}

\address{$^2$ Computational Science Center, Korea Institute of Science and Technology, Seoul 136-791, Korea}

\address{$^3$ Departamento de Matem\'aticas and Grupo Interdisciplinar de Sistemas Complejos (GISC), Universidad Carlos III de Madrid, Avenida de la Universidad 30, E-28911 Legan\'es, Spain}

\begin{abstract}
Highly ordered one-dimensional arrays of nanodots, or nanobeads, are
fabricated by forming nanoripples and nanodots in sequence, entirely by
ion-beam-sputtering (IBS) of Au(001). This demonstrates the
capability of IBS for the fabrication of sophisticated
nanostructures via hierarchical self-assembly. The intricate
nanobead pattern ideally serves to identify the governing mechanisms
for the pattern formation: Non-linear effects, especially local
redeposition and surface-confined transport, are essential both for
the formation and the preservation of the pattern order.
\end{abstract}

\pacs{81.16.Rf, 79.20.Rf}


\section{Introduction}

 Ion-beam-sputtering (IBS) has stimulated extensive experimental and theoretical studies
 due to its potential to fabricate ordered nanopatterns on many different surfaces
 in a self-organized fashion \cite{ChasonReview,munoz-garcia}.
 IBS singles out from other methods of self-assembly
 that exploit selective bonding characteristics of the assembled materials
 such as Langmuir-Blodgett films and block co-polymers,
 because it initiates the self-assembly via physical processes
 such as physical erosion of the substrate, redeposition and diffusion of ad-species.
 As a consequence, nanopatterning by IBS can be universally applied
 to materials ranging from metals \cite{Metal_Valbusa},
 semiconductors \cite{Semiconductor_Facsko}, oxides \cite{MgO}
 to polymers \cite{polymer_wei,Moon}. Thus patterned surfaces show novel catalytic \cite{Rocca},
magnetic \cite{Magnetism,Magnetism2}, and optical properties \cite{Everts},
facilitating the functionalization of various surfaces.

However, the patterns produced by IBS are essentially restricted to periodic ripples and nano-hole/dot arrays,
that are generated respectively by oblique and normal beam incidence onto the target surface.
With a few exceptions such as IBS with a rotating substrate \cite{Frost} and with an angularly dispersed ion-beam \cite{Ziberi}, most sputter-induced patterns have been fabricated simply by a single ion-beam in a single sputter geometry. Such a practice seems to limit the diversity of the patterns produced. To overcome this limitation, recently multiple-ion-beam-sputtering has been proposed to produce diverse arrangements via interference of patterns formed by several ion-beams. Within one approach, more than two ion-beams are projected onto a surface simultaneously from different orientations \cite{Carter1,Carter2,Linz1}.
Actually, some of us demonstrated that square symmetric patterns of nanoholes and nanodots can be fabricated on Au(001)
by dual-ion-beam-sputtering even at oblique incidence \cite{Joe2007}.
Another approach is sequential-ion-beam-sputtering (SIBS) of a surface, where different ion-beams are employed sequentially, changing their orientation with respect to the substrate \cite{Linz1}.
To examine this possibility, Kim {\it et al.} \cite{JHKim2009} first formed a ripple pattern in one direction by sputtering at a grazing angle, and sputtered the surface subsequently changing only the azimuthal angle by 90$^{\circ}$.
The resulting pattern is, however, not the one made by the mere superposition of the crossing ripples, which contradicts previous theoretical predictions \cite{Linz1}.

In the present work, we present yet another realization of SIBS that is aimed
at diversifying the available nanopatterns in a controlled manner.
We first fabricate a ripple pattern on Au(001) by IBS at an oblique incidence angle, and then sputter that rippled surface at normal incidence. Highly ordered nanodots form selectively on the pre-patterned ripples. We call this salient one-dimensional (1-D) feature, nanobead pattern. Again, this pattern negates the idea of solid phase superposition of patterns formed by each beam \cite{Linz1}. Instead, the nanobead pattern demonstrates the potential of SIBS to fabricate sophisticated ordered nanostructures by sequential fabrication of simple structures, in a so called hierarchical self-assembly. Note that, up to now, hierarchical self-assembly by IBS had been always guided by ordered templates previously pre-patterned by top-down approaches such as lithography \cite{GuidedSA1} and focused ion-beams \cite{GuidedSA2}. The present work proceeds, instead, via a fully bottom-up approach. This novel scheme for hierarchical self-assembly might thus overcome well-known limits imposed by the top-down fabrication of templates such as high cost, low processing speed, and limited template material and patterned area.

To understand how the nanobead patterns develop by SIBS, we have performed an extensive numerical study
that is based on the models that are to date best established in the context of nanostructuring by IBS \cite{munoz-garcia}. 
The nanobead structure serves as a test bed for model assessment, since for each model the ensuing pattern depends sensitively on the underlying physical processes that are considered, like curvature dependent sputtering yield, irradiation induced transport, surface diffusion, etc.
Agreement between experiments and numerical simulations suggests that non-linear effects, especially local redeposition effects,
play a crucial role in the evolution of the nanobead pattern. Actually, the significant role of redeposition has been already observed for highly corrugated surfaces \cite{TCKim2006,Joe2007,JHKim2009}.


\section{Experiment}

\begin{figure*}
\includegraphics[angle=0,width=0.95\textwidth]{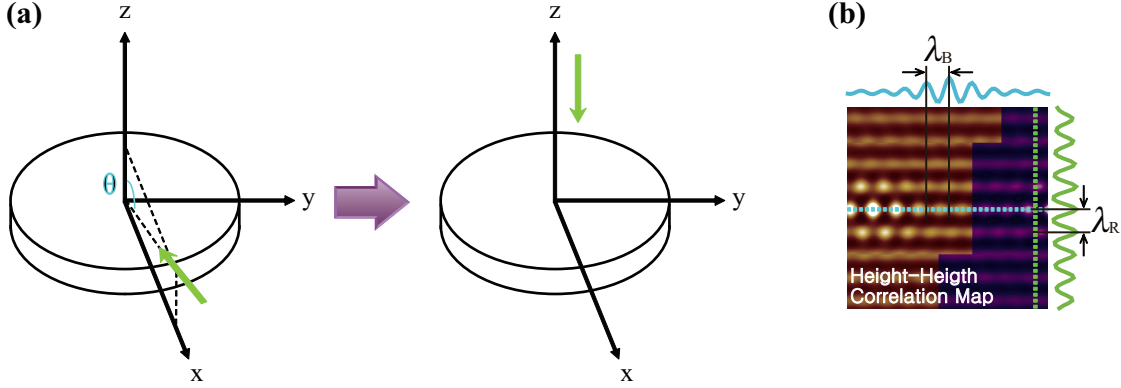}
\caption{(a) Schematic illustration of the experimental geometry for the fabrication of nanobead patterns by sequential-ion-beam-sputtering (SIBS) on Au(001) substrate. (b) From the 2-D height-height correlation map, the mean wavelength of ripple $\lambda _{\rm R}$ and mean wavelength of bead $\lambda _{\rm B}$ are retrieved.} \label{Fig. 1}
\end{figure*}

Figure~\ref{Fig. 1}(a) shows a schematic of the fabrication of nanobead patterns on Au(001) by two steps in sequence.
In the first step, ripple patterns are produced by Ar$^+$ beam sputtering of Au(001) along the densely packed [110] direction with a polar angle $\theta$ = 72$^{\circ}$ from the surface normal.
After fabricating the initial ripple patterns, we sputter further each rippled surface at normal incidence.

For the formation of initial ripple patterns, the partial pressure of Ar$^+$, $P_{Ar}$, the ion energy $\varepsilon$, the ion flux $f$, and the ion fluence $\psi $ were 1.2 $ \times$ 10$^{-4 }$ Torr, 2 keV, 0.3 ions nm$^{-2}$ s$^{-1}$, and 4500 ions nm$^{-2}$, respectively.
The ion fluence is defined as the ion flux multiplied by the accumulated sputter time.
For the subsequent sputtering at normal incidence, $P_{Ar}$, $\varepsilon $, and $f$ were 1.2 $ \times$ 10$^{-4 }$ Torr, 2 keV, and 1.1875 ions nm$^{-2}$ s$^{-1}$, respectively.
All the experiments for the sample sputtering were performed in an ultrahigh vacuum chamber with a base pressure of about 5 $\times$ 10$^ {-10}$ Torr.
During the sputtering, the sample temperature was kept around 300 K.
The patterned surface was then analyzed \textit{ex situ} by an atomic force microscope (AFM) in the contact mode and a scanning electron microscope (SEM).

From the AFM images of the nanobead pattern, we retrieve the following structural information: The surface roughness $W$ is defined as $W(t)\equiv\sqrt{\langle[h(\textbf{r},t)-\overline{h}(t)]^{2}\rangle}$, where $\overline{h}(t)$  is the mean height at time $t$.  The mean ripple wavelength $\lambda _{\rm R}$(t) and mean nanobead wavelength $\lambda _{\rm B}$(t) are obtained from the two-dimensional (2-D) height-height correlation function $G(\textbf{r})$ of AFM images as illustrated in Fig.~\ref{Fig. 1}(b). Here, $G(\textbf{r})$ is defined as $G(\textbf{r})\equiv\langle h(\textbf{r}+\textbf{r}_{\rm i}) h(\textbf{r}_{\rm i})\rangle$,
with $\lambda _{\rm R}$ being estimated as the distance between the central ripple and the neighboring ones from a 2-D map of $G(\textbf{r})$. Finally, $\lambda _{\rm B}$ is obtained from the line profiles along the central ripples of the 2-D map of $G(\textbf{r})$. Due to the compact arrangement of beads along each ripple, $\lambda _{\rm B}$ can be interpreted as the mean diameter $\overline{D}$ of nanobeads. The mean nanobead amplitude $A_{B}$ is estimated from the line profiles taken along the centers of ripples from numerous images.

\section{Result}

\begin{figure*}
\includegraphics[angle=0,width=1\textwidth]{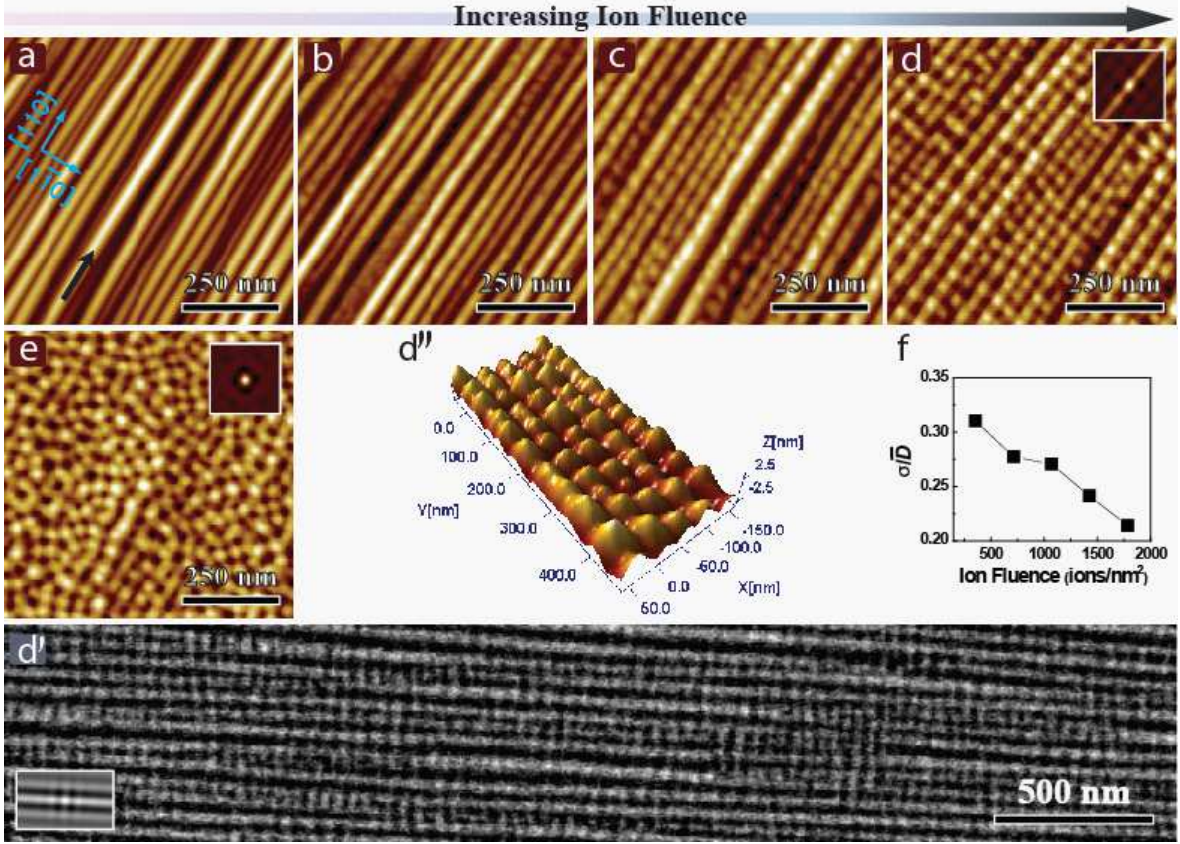}
\caption{(color online) (a-e) AFM images. (a) Initial, rippled surface with $\lambda _{\rm
R} \simeq$ 47 nm and $W \simeq$ 2.7 nm. The arrow indicates the
incidence direction of the ion-beam as projected onto the target
plane. Surface morphologies after sputtering normal to the rippled
surface with (b) $\psi $ = 356 ions nm$^{-2}$, (c) $\psi $ = 1069
ions nm$^{-2}$ and (d) $\psi $ = 1781 ions nm$^{-2}$. (e) Surface
morphology after sputtering normal to an initially flat surface with
$\psi $ = 1781 ions nm$^{-2}$. The insets are 2-D height-height
correlation maps. (d$^{\prime}$) SEM image of the sample giving the
image (d). (d$^{\prime\prime}$) 3-D image of ordered nanobead pattern
of the sample (d). (f) Evolution of $\sigma$/$\overline{D}$, where
$\sigma$ is the standard deviation of the distribution of the
nanobead diameter $D$.
 Image size: (a-e) 750 $\times$ 750 nm$^{2}$ and (d$^{\prime}$) 3600 $\times$ 680 nm$^{2}$.}
\label{Fig. 2}
\end{figure*}

Figure~\ref{Fig. 2}(a) shows a typical ripple pattern formed after sputtering at an oblique angle $\theta$ = 72$^{\circ}$.
For this structure, $\lambda _{\rm R}$ and $W$ are 47 nm and 2.7 nm, respectively.
The ripples form along the direction of the incident ion-beam, that is chosen to coincide with the densely packed crystallographic direction [110]. Figs.~\ref{Fig. 2}(b-d) illustrate the development of the nanobead patterns upon the initial ripples by subsequent ion-beam-sputtering at normal incidence with increasing ion fluence $\psi$.

At an ion fluence $\psi $ = 356 ions nm$^{-2}$ [Fig.~\ref{Fig. 2}(b)],
we can observe nanobeads aligned along the initial ripple, with $\lambda _{\rm B}\simeq$ 35 nm and $A_{\rm B}\simeq$ 0.5 nm. However, the mean coherence length of the nanobead $\ell_{\rm B}$ is less than 240 nm, equivalent to the mean length of seven beads. The coherence length is defined as the mean length of a series of beads that are aligned without interruption along the ripple with a well-defined period.

At an ion fluence $\psi $ = 1069 ions nm$^{-2}$, nanobeads are found to cover most of the top area of the ripples [Fig.~\ref{Fig. 2}(c)]. Now $\lambda_{\rm B}$,  $A_{\rm B}$, and $\ell_{\rm B}$ increase to 46 nm, 0.7 nm, and 600 nm, respectively. With further increase of ion fluence $\psi $ = 1781 ions nm$^{-2}$, we can observe {\it well-ordered} nanobeads over the whole top area [Fig.~\ref{Fig. 2}(d)]. Each bead has grown further with $\lambda_{ \rm B} \simeq$ 50 nm, $A_{\rm B} \simeq$ 1.5 nm, and  $\ell_{\rm B} > 1$ $\mu$m that is the limit of our AFM. Figure~\ref{Fig. 2}(d$^{\prime}$) is the image of the same sample taken by a SEM that gives a wider areal view. The image shows the nanobead pattern to extend over the whole ripple length, implying  $\ell_{\rm B} > 3$ $\mu$m.

Figure~\ref{Fig. 2}(f) shows that the standard deviation of the nanobead size distribution, normalized by the mean bead diameter or mean bead wavelength, $\sigma$/$\overline{D}$, also decreases with increasing $\psi$,
from 0.31 for Fig.~\ref{Fig. 2}(a) down to a minimal value, 0.21 for Fig.~\ref{Fig. 2}(d).
This indicates that as sputtering proceeds, the nanobead size becomes more and more uniform.

The height-height correlation functions $G(\textbf{r})$ of the nanobead patterns of Figs.~\ref{Fig. 2}(d) and 2(d$^{\prime}$) are displayed in the inset of each figure. A square symmetric pattern is observed around the central peak, implying that the nanobeads are well-ordered not only along the ripple direction [110], but also along the inter-ripple direction [1$\overline 1 $0], although the inter-ripple correlation of the beads is weak relative to the intra-ripple correlation. Such an order between the beads on adjacent ripples indicates that adatoms generated by the normal incidence irradiation can efficiently diffuse along the close packed [1$\overline{1}$0] direction that is perpendicular to the ripples, and mediate the correlated growth of the beads in the neighboring ripples. Figure~\ref{Fig. 2}(d$^{\prime\prime}$) displays a nanobead pattern, clearly showing its 1-D nature in a 3-D perspective.

The well-defined order of the 1-D nanobead pattern of Fig.~\ref{Fig. 2}(d) contrasts strikingly with the relatively poor order of the 2-D nanodot pattern of Fig.~\ref{Fig. 2}(e) that forms on an initially flat surface under the same sputtering condition that induce the nanobead pattern on the initially rippled surface. Thus, the ordered growth of the latter should originate from the initial condition imposed by the pre-patterned ripples, which limit the kinetic processes randomizing the growth of nanodots
and guide their growth along the initial ripples.

Note that Lian {\it et al}.\ \cite{GuidedSA2} have also reported bead
patterns formed by focused-ion-beam irradiation of Co strips and
rings formed on silicon oxides, with the lines of beads increasing
for wider Co templates. However, in these cases, the bead formation
is triggered by dewetting of the Co strips or rings to minimize
surface free energy, as a manifestation of the Rayleigh
instability \cite{Rayleigh}. In this case, a well-defined relation
between the width of the strip or ring and the period of beads,
$\lambda_{\rm B}$ $>$ $\pi$ $\lambda_{\rm R}$, is predicted to occur
and was actually observed in their experiments. However, such a
relation is not met at all for the present case, and the present
nanobead formation is, thus, not driven by the Rayleigh instability.

\section{Discussion}

In general, nano-scale pattern formation by IBS has been explained theoretically by continuum models with different levels of sophistication. Note that for the present flux conditions nontrivial morphology changes occur in macroscopic time scales (of the order of seconds) that are not accessible to more atomistic approaches. A linear model based on Sigmund's theory \cite{Sigmund1} 
of ion erosion combined with Mullins's surface-diffusion theory \cite{Mullins} 
was proposed by Bradley and Harper (BH), and could elucidate formation of ripples and their orientation with respect to the ion-beam \cite{BH}. Other features like ripple stabilization are, however, beyond the capabilities of the BH
model. To overcome such shortcomings, various nonlinear generalizations have been introduced
like the Kuramoto-Sivashinsky (KS) model \cite{MakeevReview} 
or the extended KS (eKS) model \cite{Kim}, that were able to reproduce features such as
onset of kinetic roughening or ripple coarsening. Besides, the damped KS model \cite{Facsko} 
has been also suggested to reproduce pattern formation by IBS, although its physical interpretation 
remains unclear.

Following an alternative route, a ``hydrodynamic'' approach \cite{HydroDynModel} to IBS has also led to the eKS model, and could identify local redeposition of sputtered material as the major physical effect behind the improved description of IBS provided by this model. Although the first principles description of IBS is a subject of current debate (see several related papers e.g.\ in \cite{jpcm}), to date the eKS equation provides a rather complete qualitative description of IBS, that has recently been shown to describe quantitatively experiments on Si targets in a self-consistent fashion \cite{Javier2010}, and can be expected to apply also to metallic targets. Thus, within this ``hydrodynamical" formulation \cite{HydroDynModel,munoz-garcia}, the dynamics is described both for the surface height $h(\mathbf{r},t)$ of the target, and for the density $R(\mathbf{r},t)$ of species (for metals, e.g.\ adatoms, advacancies \cite{ChasonReview,Metal_Valbusa}) that are subject to transport at the surface and can locally redeposit back to the immobile bulk. In principle, one thus needs to solve a system of two coupled evolution equations. However, due to the smallness of the rate of ion arrival as compared to the rate of atomistic relaxation processes (e.g.\ surface diffusion hopping attempts), one can solve approximately the time evolution of $R(\mathbf{r},t)$ and introduce the result into the equation for $h(\mathbf{r},t)$, that finally reads \cite{munoz-garcia}
\begin{equation}
\frac{\partial h}{\partial t} = - \nu \nabla^{2} h  - {\cal K} \nabla^{4} h + \lambda^{(1)} (\nabla h)^{2}  + \lambda^{(2)} \nabla^{2}(\nabla h)^{2} .
\label{eq0}
\end{equation}
In this equation, coefficients $\nu$, ${\cal K}$, $\lambda^{(1)}$, and $\lambda^{(2)}$ are functions of phenomenological parameters like the average ion energy and flux, temperature, etc., see Ref.\ \cite{munoz-garcia} and references therein. In particular, local redeposition induces a non-zero value of coefficient $\lambda^{(2)}$, and contributes additionally to coefficient ${\cal K}$. Eq.\ (\ref{eq0}) is the eKS model which, after suitable rescaling of $h$, $\mathbf{r}$ and $t$, can be expressed as \cite{munoz-garcia}
\begin{equation}
\frac{\partial h}{\partial t} = - \nabla^{2} h  - \nabla^{4} h + (\nabla h)^{2}  + r \nabla^{2}(\nabla h)^{2} + \eta ,
\label{eq1}
\end{equation}
where $r=(\nu \lambda^{(2)})/({\cal K} \lambda^{(1)})$ can be interpreted as the ratio of the (squared) nonlinear crossover length scale $|\lambda^{(2)}/\lambda^{(1)}|$ to the (squared) linear crossover length scale ${\cal K}/\nu$, and takes negative values for physical conditions \cite{HydroDynModel}. In general, larger $r$ values correspond to conditions under which local redeposition is enhanced leading to stronger coarsening and local ordering properties \cite{munoz-garcia}. The advantage of studying the rescaled equation (\ref{eq1}) is that it allows straightforward analysis of the system as function of the single parameter $r$ on which qualitative properties are thus seen to depend. We have additionally introduced noise fluctuations in Eq.\ (\ref{eq1}) to account for the randomness of ion arrival, herein described by an uncorrelated, zero-mean white noise $\eta$.

Eq.\ (\ref{eq1}) contains both the BH and the KS models as particular cases that are obtained, respectively, by neglecting the two nonlinear terms, or by simply setting $r = 0$. In comparison with these two models, the distinctive feature of the eKS model is the additional presence of the so-called conserved KPZ (cKPZ) nonlinear term with parameter $r$. As mentioned above, physically it reflects \cite{munoz-garcia,HydroDynModel} surface-confined transport of species that have been dislodged from the crystalline target but remain on the target surface (local redeposition). Although for semiconductors this mechanism is currently under debate in competition with other relaxation mechanisms such as viscous flow, stress, etc.\ \cite{jpcm} we expect it to be a relevant transport mechanism for metallic surfaces, albeit within a simplified description in which anisotropies to surface diffusion are neglected.

In order to understand how the nanobead patterns develop and evolve by IBS of rippled surfaces, we perform extensive numerical studies of the relevant models such as the BH, KS, and eKS, trying to assess in particular the relevance of redeposition. Our numerical integration employs centered differences for spatial derivatives, Euler method for time evolution, and Lam and Shin's \cite{Shin} discretization for the nonlinear terms.

\begin{figure*}
\includegraphics[angle=0,width=1\textwidth]{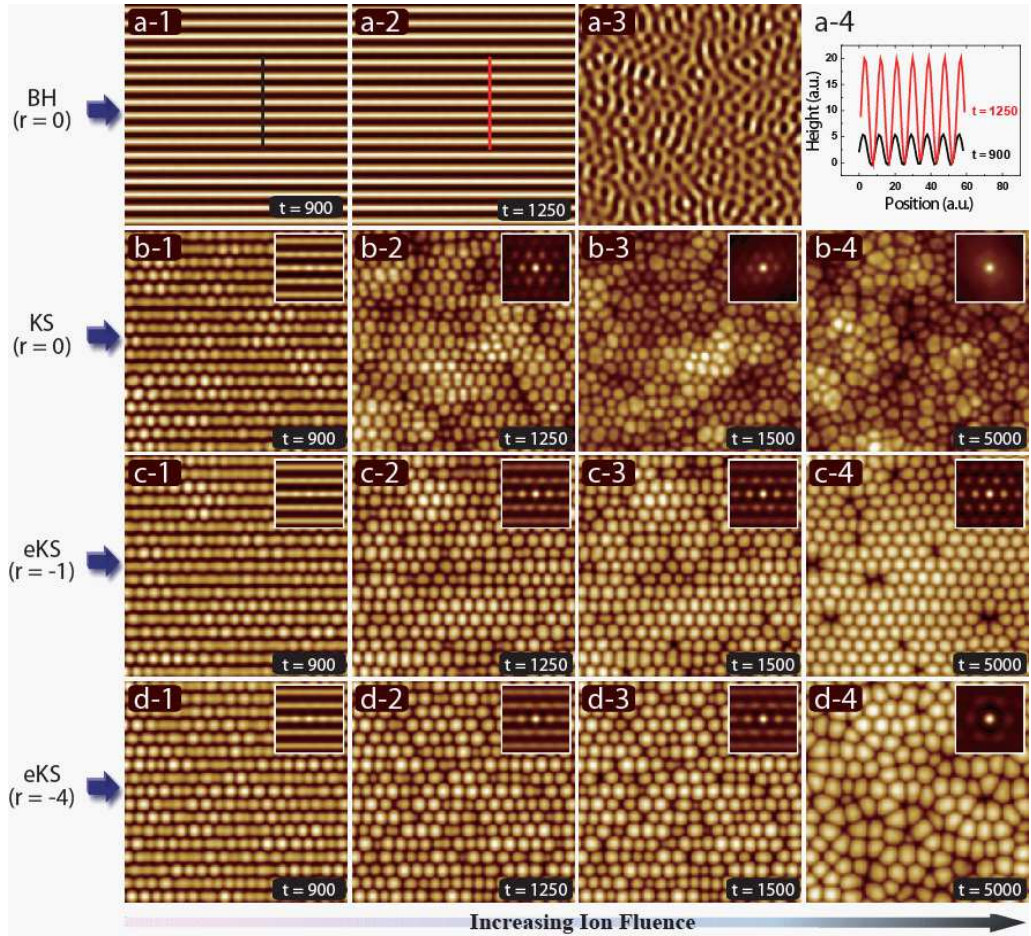}
\caption{(color online) Simulated images at representative simulation times for the (a) BH, (b) KS ($r$ = 0), (c) eKS ($r = -1$), and (d) eKS ($r = -4$) models. (a-3) is obtained by differentiating the height of (a-2) along the ripple direction. (a-4) shows the height-profiles along the two lines marked respectively in (a-1) and (a-2). Images on the same column are taken at the same simulation times. The simulation time corresponding to each column is also marked with arrows labeled by each column number, from 1 to 4 in Fig.~\ref{Fig. 4}(a). For  the KS ($r$ = 0) and the eKS ($r = -1$, $r = -4$) models, the images on the left-most column show almost maximal order. } \label{Fig. 3}
\end{figure*}

\begin{figure*}
\includegraphics[angle=0,width=1\textwidth]{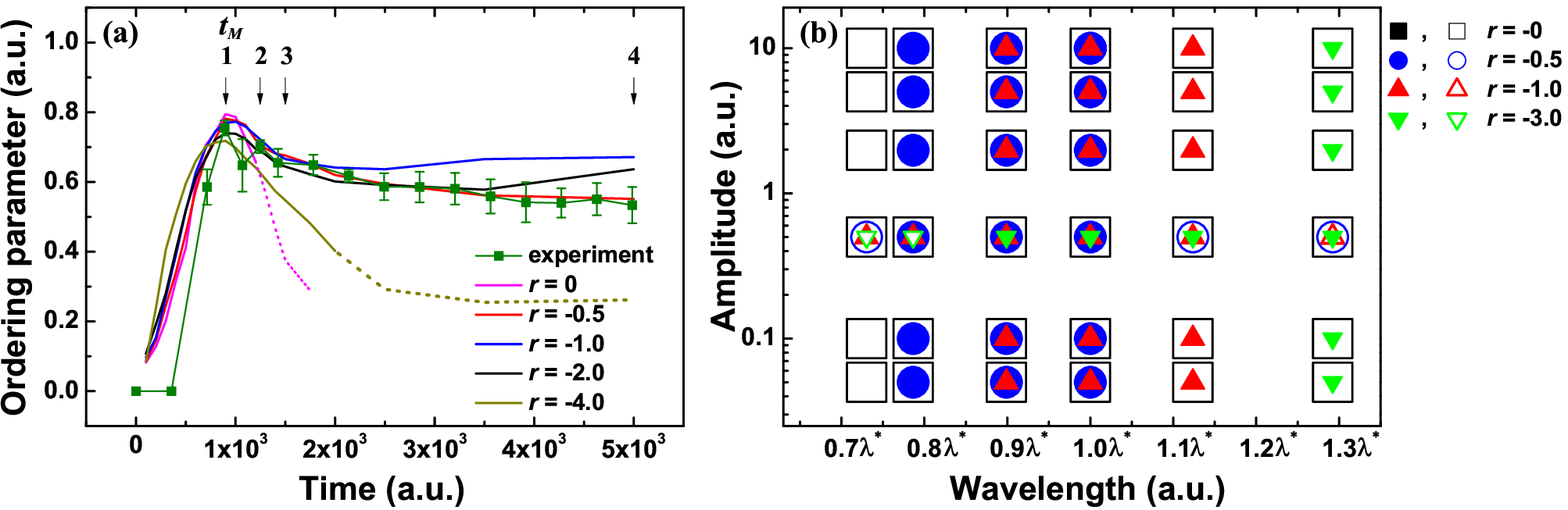}
\caption{(color online) (a) Temporal evolution of the ordering parameter for the experimental and simulated nanobead patterns for $r$ values
from 0 to $-4$. The relative intensity of the first order peak to that of the zeroth order peak along the central ripple on the 2-D
height-height correlation map is taken as the ordering parameter.
Solid (dashed) lines denote the case where well-defined nanobead (poor or 2-D like) pattern forms.
Simulation time is scaled to the experimental ion fluence so as to synchronize the time showing the maximum ordering parameter.
Each arrow on the upper side of the figure that is labeled by a number denotes the simulation time of the images in the corresponding numbered columns in Fig.~\ref{Fig. 3}.
(b) The temporal evolution of the ordering parameter shows different behavior depending on the wavelength and amplitude
of the initial ripple pattern and $r$. Solid (open) symbols indicate the case where the ordering parameter preserves more
than a half of its maximum value at $t_M$ (decrease very steeply within $t_M$) after reaching its maximum. Here,
$t_M$ is the time for the ordering parameter to reach its maximum value, after sputtering starts normal to the rippled surface.}
\label{Fig. 4}
\end{figure*}

Figure~\ref{Fig. 3} shows the simulated surface morphologies after IBS normal to the  pre-rippled surface according to the (a) BH, (b) KS, (c) and (d) eKS models with $r = -1$ and $r = -4$, respectively. For all cases, the initial ripple wavelengths are chosen to be $\lambda^{*}$, that is, the wavelength of ripples predicted by the linear BH instability.

Figures~\ref{Fig. 3}(a-1) and ~\ref{Fig. 3}(a-2) are simulated images at two different sputter times according to the BH model, in which no ordered nanobead pattern is observed. Instead, some irregular pattern inscribed on the high rising ripples is revealed in Fig.~\ref{Fig. 3}(a-3), that is obtained by differentiating the height of Fig.~\ref{Fig. 3}(a-2) in the ripple direction, along which the height varies little. Figure~\ref{Fig. 3}(a-4) shows profiles along the lines on Figs.~\ref{Fig. 3}(a-1) and ~\ref{Fig. 3}(a-2), indicating that just the amplitude of the ripples increases.

For both the KS and eKS models, on the other hand, well-defined nanobead patterns develop by normal IBS of the pre-rippled surfaces. Figure~\ref{Fig. 3}(b-1) shows an optimal nanobead pattern for the KS model, and Figs.~\ref{Fig. 3}(c-1) and ~\ref{Fig. 3}(d-1) for the eKS model with $r = -1$ and $-4$, respectively. The height-height correlation maps in their insets show predominantly 1-D order of dots, although weak,  short-range hexagonal order with the beads in the neighboring ripples is also observed.
This implies that nonlinear effects 
are essential for the nanobead formation. Note that in our simulations, the substrate is assumed to be amorphous, while the drive towards close-packing of dots dictates their hexagonal order \cite{Facsko_hex,HydroDynModel}. In our experiments, however, the nanobead patterns reveal a square symmetric order (Fig.~\ref{Fig. 2}). This suggests that the anisotropic diffusion of ad-species via the efficient channels along \{110\} directions of crystalline Au(001) plays a significant role for the development of order between the beads in neighboring ripples.

There is, however, a noticeable difference in the temporal evolution of the order upon extended ion-beam-sputtering between the KS and the eKS models, namely, further sputtering makes the beads overgrow ripples, and their inter-ripple correlation becomes apparent. [See Figs.~\ref{Fig. 3}(b-2),~\ref{Fig. 3}(c-2), and ~\ref{Fig. 3}(d-2)] Still, the 1-D order of beads is preserved as shown by the strong intensities of the dots in the central ripple compared with those of the neighboring ripples in the height-height correlation map of each figure.
As sputtering proceeds further, the spot intensities in the height-height correlation maps in Figs.~\ref{Fig. 3}(b-3), ~\ref{Fig. 3}(c-3), and ~\ref{Fig. 3}(d-3) becomes weak with respect to the central spot, indicating that the spatial order of beads becomes weak. For the KS model, 1-D order is already obscure in Fig.~\ref{Fig. 3}(b-3), while it is still well discernable for the eKS model. With further sputtering, the spatial order of dots becomes poorer. [See Figs.~\ref{Fig. 3}(b-4),~\ref{Fig. 3}(c-4), and ~\ref{Fig. 3}(d-4)] Even no orientational order of dots is observed for the KS model as seen in the height-height correlation map in Fig.~\ref{Fig. 3}(b-4). For the eKS model, the temporal evolution of order changes, depending on the relative strength of $r$.  With $r = -4$, the order of dots looks 2-D like as noticed from the inset of Fig.~\ref{Fig. 3}(d-4), while for $r = -1$, 1-D order of dots is still well preserved. Thus, there seems to be an optimal strength ($r$ value) of the cKPZ term for nanobead pattern enhancement.

In order to perform a semi-quantitative analysis of the temporal evolution of the order in the nanobead pattern, we choose to define an ordering parameter as the intensity of the first order peak relative to that of the zeroth order peak along the central ripple of the height-height correlation map of the nanobead pattern. In Fig.~\ref{Fig. 4}(a), the temporal evolution of the ordering parameter is displayed for the simulated nanobead patterns for various $r$ from 0 to $-4$,
along with the experimental ones. The wavelength of  pre-patterned ripple is chosen to be $\lambda ^{*}$.
For all the simulations, the ordering parameters show maxima around $t_M$ = 900 in simulation units.
As noticed in Fig.~\ref{Fig. 3}, there is a difference in the temporal evolution of the ordering parameter between the KS and the eKS models. For the KS model ($r$ = 0), the ordering parameter degrades very rapidly, as compared with the cases in which $r$ is not zero (eKS model).
For $r = -0.5, -1$, and $-2$, 1-D order of the nanobead pattern is preserved for quite a long time.
For $r = -4$, however, order is lost at a relatively early time. We have examined the temporal evolution of the ordering parameter for a wide range of initial ripple amplitudes ($W$) spanning three orders of magnitude, wavelength values ($\lambda$) around $\lambda^*$, and several values of $r$; results are summarized in Fig.~\ref{Fig. 4}(b).
In the figure, solid (open) symbols denote cases in which the ordering parameter decreases relatively slowly or
preserves its value for much longer times than the time required to reach the maximal order parameter, $t_M$ $\simeq$ 900 simulation units (decay substantially in such a time interval as $t_M$), after reaching maximum. Notably, the KS model predicts substantial decay of the ordering parameter within $t_M$ after reaching its maximum value, irrespective of the wavelength and amplitude of initial ripple. This strongly indicates that the cKPZ term is crucial for order stabilization of the nanobead pattern.

The experimentally observed ordering parameter is also displayed in Fig.~\ref{Fig. 4}(a) after rescaling the experimental time and magnitude of the ordering parameter, so that it shows its maximum at the same time and with the same magnitude as the mean theoretical ones. After reaching maximum, the experimentally observed ordering parameter decreases slightly,
and then remains high for a long time, seemingly being at a stationary state.
All these details are well reproduced by the eKS model, especially with $r = -0.5$ in Fig.~\ref{Fig. 4}(a).
Thus, as in the case of Si targets \cite{Javier2010}, the eKS model seems a self-consistent model to reproduce the intricate temporal evolution of nanobead pattern, even at a semi-quantitative level.

As mentioned above, the cKPZ nonlinearity appearing in the eKS model represents local redeposition effects \cite{HydroDynModel} that appear to be essential in order to heal the ion-eroded surface, seeming to preserve the ordering parameter as observed in Fig.~\ref{Fig. 4}(a). In analogy with pattern formation in macroscopic systems such as ripples on sand dunes, surface confined transport of redeposited material tends to promote lateral coarsening of pattern features, which is more pronounced for large $r$ values \cite{munoz-garcia}. In our case, this effect competes with alignment of the nanobeads along the original ripples on the target, there being an optimal balance at $r=-0.5$ between both trends that provides an arrangement with the best ordering. For instance, for $r = -4$ in Fig.~\ref{Fig. 3}(d-4) it is apparent that coarsening of individual beads up to a larger stationary lateral size hinders 1-D bead alignment.



\section{Summary and Conclusions}

We have fabricated a salient nanobead pattern on Au(001) by sequential-ion-beam-sputtering. This demonstrates the capability of IBS for hierarchical self-assembly of sophisticated nanostructures that is moreover achieved via a fully bottom-up approach. Considering the universal character of nanopatterning by IBS, the present scheme of hierarchical self-assembly should be transferable to most other materials, and is expected to open a new avenue for this technique. The eKS model self-consistently reproduces most of the experimental details for the formation of the nanobead patterns in contrast with other available models, and thus allows to conclude on the importance of local redeposition and surface-confined transport in these nanoscopic pattern formation systems. Improvements in the quantitative continuum description of the present experiments might be expected from a more detailed description that takes into account anisotropies in surface diffusion that are typical of metallic substrates.


\ack
The work was supported by NRF (Korea) Grant No.\ 20100010481 and by MICINN (Spain) Grant No.\
FIS2009-12964-C05-01. R.\ C.\ gratefully acknowledges warm hospitality at Sook-Myung Women's University while part of this work was done.

\end{document}